\documentclass[twocolumn]{aastex63}

\newcommand{\lrb}[1]{\left( #1 \right)}
\newcommand{\lrbs}[1]{\left[ #1 \right]}
\newcommand{\cory}[1]{{\color{black}{#1}}}
\newcommand{\coryfr}[1]{{\color{black}{#1}}}
\received{}
\revised{}
\accepted{April 21 2021}

\submitjournal{ApJ}

\shorttitle{Dynamical Formation of Radio sources in Globular Clusters}
\shortauthors{Zhao et al.}

\graphicspath{{./}{figures/}}

\begin{document}

\title{The MAVERIC Survey: Dynamical Origin of Radio Sources in Galactic Globular Clusters}

% Author List
\author[0000-0002-9547-8677]{Yue Zhao}
\affiliation{Department of Physics, 
University of Alberta, \\
Edmonton, AB T6G 2E1, Canada}

\author[0000-0003-3944-6109]{Craig O. Heinke}
\affiliation{Department of Physics, 
University of Alberta, \\
Edmonton, AB T6G 2E1, Canada}

\author[0000-0003-0286-7858]{Laura Shishkovsky}
\affiliation{Center for Data Intensive and Time Domain Astronomy, \\
Department of Physics and Astronomy, Michigan State University, \\
East Lansing, Michigan 48824, USA}

\author[0000-0002-1468-9668]{Jay Strader}
\affiliation{Center for Data Intensive and Time Domain Astronomy, \\
Department of Physics and Astronomy, Michigan State University, \\
East Lansing, Michigan 48824, USA}

\author[0000-0002-8400-3705]{Laura Chomiuk}
\affiliation{Center for Data Intensive and Time Domain Astronomy, \\
Department of Physics and Astronomy, Michigan State University, \\
East Lansing, Michigan 48824, USA}

\author[0000-0003-0976-4755]{Thomas J. Maccarone}
\affiliation{Department of Physics and Astronomy, Texas Tech University, \\
Box 41051, Lubbock, TX 79409-1051, USA}

\author[0000-0003-2506-6041]{Arash Bahramian}
\affiliation{International Centre for Radio Astronomy Research, \\
Curtin University, GPO Box U1987, Perth, WA 6845, Australia}

\author[0000-0001-6682-916X]{Gregory R. Sivakoff}
\affiliation{Department of Physics, 
University of Alberta, \\
Edmonton, AB T6G 2E1, Canada}

\author[0000-0003-3124-2814]{James C. A. Miller-Jones}
\affiliation{International Centre for Radio Astronomy Research, \\
Curtin University, GPO Box U1987, Perth, WA 6845, Australia}

\author[0000-0002-4039-6703]{Evangelia Tremou}
\affiliation{LESIA, Observatoire de Paris, CNRS, PSL, SU/UPD, Meudon, France}

\begin{abstract}
We investigate potential correlations between radio source counts (after background corrections)  
of 22 Galactic globular clusters (GCs) from the MAVERIC survey, and stellar encounter rates ($\Gamma$) and masses ($M$) of the GCs. 
Applying a radio luminosity limit of $L_\mathrm{lim} = 5.0 \times 10^{27}~\mathrm{erg~s^{-1}}$, we take a census of radio sources in the core and those within the half-light radius, of each cluster. By following a maximum likelihood method and adopting a simplified linear model, we find an unambiguous dependence of core radio source counts on $\Gamma$ and/or $M$ at $90\%$ confidence, but no clear dependence of source counts within the half-light radius on either $\Gamma$ or $M$. 
Five of the identified radio sources in GC cores above our adopted limit are millisecond pulsars or neutron star X-ray binaries (XRBs), the dependence of which on $\Gamma$ is well-known, but another is a published black hole (BH) XRB candidate, and ten others are not identified; \coryfr{Accounting for these verified cluster members increases the significance of correlation with $M$ and/or $\Gamma$ (to 99\% confidence), for fits to core and half-light region source counts, while excluding a dependence on $\Gamma$ alone at 90\% (core) and 68\% (half-light) confidence}. 
This is consistent with published dynamical simulations of GC BH interactions that argue $\Gamma$ will be a poor predictor of the distribution of accreting BHs in GCs.
 %\cory{The former is more significant---at $99\%$ confidence---when we consider confirmed members in the clusters.} 
%Our work indicates that radio sources in GC cores may be produced dynamically or primordially, likely by a combination of both routes. 
Future multiwavelength follow-up to verify cluster membership will enable stronger constraints on the dependence of radio source classes on cluster properties, promising a new view on the dynamics of BHs in GCs.
%It is not yet possible to constrain the dependence of sources within the half-light region, due to the preponderance of background sources, which will require multiwavelength follow-up to discriminate. 
%is less supportive for black hole or neutron star binaries as a result of mass segregation and relatively low encounter rates.
\end{abstract}

\keywords{Globular star clusters (656) --- Low-mass X-ray binary stars (939) --- Compact objects (288) --- Neutron stars (1108) --- Black holes (162) --- Stellar dynamics (1596)}

\section{Introduction}
\label{sec:introduction}
Since the launch of early X-ray missions like {\it Uhuru} and {\it OSO-7}, it has been known that Galactic globular clusters (GCs) host an overabundance of X-ray sources \citep{Katz75}. These X-ray sources are ascribed to close binaries of various kinds, whose origin is closely related to the dynamically active cores of GCs. Specifically, the very dense cores of GCs facilitate many close dynamical encounters, producing close binaries through multiple possible channels \citep{Clark75b, Fabian75, Sutantyo75, Hills76}. The most %typical
well-studied 
population is comprised of low-mass X-ray binaries (LMXBs), where neutron stars (NSs) or black holes (BHs) accrete matter from (usually) a (near) main-sequence donor star \citep{Lewin83, Grindlay84}. LMXBs were discovered to dominate the bright X-ray source population, with typical X-ray luminosities $L_X\sim 10^{36-38}~\mathrm{erg~s^{-1}}$ \citep{Giacconi74, Clark75a, Canizares75}. 

With the advent of the {\it Chandra X-ray Observatory}'s %instruments of
superior sensitivity and angular resolution, a plethora of faint ($L_X \lesssim 10^{33}~\mathrm{erg~s^{-1}}$) sources were detected in many GCs \citep[e.g.,][]{Grindlay01a,Grindlay01b,Pooley02a,Pooley02b,Bassa04,Heinke05,Heinke06,Bassa08,Lu09,Zhao19}. Some of these faint sources are thought to be quiescent LMXBs (qLMXBs), which are generally $\lesssim 10^4$ times fainter than actively accreting LMXBs \citep[e.g.,][]{Campana98, Rutledge02, Heinke03}. Closely related to LMXBs are the millisecond pulsars (MSPs); these are fast-spinning radio pulsars spun up by accreted matter from donor stars during a prior LMXB phase \citep{Bhattacharya91}. MSPs emit thermal or non-thermal X-rays, and also contribute to the faint source population, especially below $10^{31}$ erg s$^{-1}$ \citep{Saito97, Bogdanov06}. %It was also noted that a considerable fraction 
A majority of the faint X-ray population between $10^{31}<L_X<10^{32}$ erg s$^{-1}$ is comprised of cataclysmic variables (CVs), which are accreting white dwarfs in close orbit with normal stars \citep{Hertz83a, Cool95, Pooley02a, Cohn10, RiveraSandoval18}. Finally, chromospherically active binaries (ABs), which are tidally-locked close binaries of normal or evolved stars, are very common in GCs below $10^{31}$ erg s$^{-1}$ \citep{Bailyn90, Dempsey93, Grindlay01a,Bassa04,Heinke05}.  The LMXB and MSP populations, along with the brighter CVs, %do depend on how frequent
are closely correlated with the frequency of dynamical encounters,  quantified by the stellar encounter rate \citep{Verbunt87,Johnston92, Verbunt03,Pooley03, Heinke03,Pooley06,Hui10,Bahramian13,Heinke20}. 
However, other X-ray sources are not of dynamical origin; most ABs descend from primordial binaries \citep{Bassa04,Bassa08,Huang10,Cheng18}, as do many CVs \citep{Davies97,Kong06,Ivanova06,Belloni19}.

Binaries containing BHs may not be distributed in the same way as those containing NSs, as the numbers of BHs remaining in different GCs, and their level of mixture with other stars, may vary in complicated ways. Early theoretical calculations suggested that dynamical interactions involving BHs would quickly expel all, or nearly all, BHs from GCs \citep{Sigurdsson93,Kulkarni93}. However, observations of candidate BH X-ray binaries in extragalactic \citep{Maccarone07} and Galactic GCs \citep{Strader12}, along with GC simulations that left numerous BHs in clusters \citep{Morscher13,Sippel13,Heggie14,Morscher15}, changed the prevailing wisdom. There is now solid dynamical evidence of three (noninteracting) BH binaries in the GC NGC 3201 \citep{Giesers18,Giesers19}. Current GC simulations predict large numbers of BHs in clusters with relatively little mass segregation 
\citep{Peuten16,Askar18,Weatherford18,Weatherford20}, though the number of detectable BHs in binary systems with other stars is not predicted to correlate with the total number of BHs in the clusters \citep{Kremer18,ArcaSedda18,Askar18}.

The MAVERIC ({\it Milky way ATCA and VLA Exploration of Radio sources In Clusters}) survey involves deep radio imaging of 50 Galactic GCs (see \citealt{Tremou18}, \citealt{Shishkovsky20}, Sh20 hereafter, and Tudor et al. in prep.), dedicated to discovery of potential BH LMXBs and other exotic radio sources.
The key motivation is that accreting BHs are more radio-luminous for a given X-ray luminosity than other systems \citep{Maccarone05,Migliari06}, making deep radio surveys an excellent method to find candidate BHs.
%, and to probe the dynamical properties of GCs. 
Thanks to the superior sensitivity and  sub-arcsecond resolving power of the {\it Karl G. Jansky Very Large Array} ({\it VLA}) and the {\it Australia Telescope Compact Array} ({\it ATCA}), it has led to fruitful revelations of faint radio sources that are strong candidate BH LMXBs \citep{Strader12, Chomiuk13, Miller-Jones15,Bahramian17,Shishkovsky18,Bahramian20}, and strong candidates for MSPs not yet detected via radio pulsations  \citep{Bahramian18,Zhao20,Urquhart20}. The $5~\sigma$ detection limit of the observations reaches as low as $S_\nu \sim 10~\mathrm{\mu Jy~beam^{-1}}$ ($S_\nu$ is the flux density), covering different kinds of radio emitting objects in the fields of the GCs. 

Of the detected radio sources, we expect a considerable fraction to be background sources (e.g., active galactic nuclei), while the sources that are actually associated with the clusters are mostly BH or NS LMXBs and MSPs. %Indeed, 
Radio emission has been observed in both NS \citep{Migliari06} and BH \citep{Gallo14, Gallo18} LMXBs, characterized by flat ($-0.5<\alpha<0$) \footnote{$\alpha$ is the radio spectral index ($\alpha$) defined by $S_\nu \propto \nu^\alpha$} to inverted ($\alpha>0$) radio spectra, while MSPs are generally steep-spectrum radio sources \citep[][$\alpha\approx -1.4$ with unit standard deviation]{Bates13}. CVs and ABs may also be active radio emitters. For example, radio emission has been observed from both non-magnetic \citep[e.g.,][]{Coppejans15, Coppejans16} and magnetic CVs \citep[e.g.,][]{Barrett17}, generally fainter than from LMXBs, but visible at GC distances (typically a few kpc's) during the peaks of outbursts \citep[e.g.,][]{Mooley17}. ABs have also been observed in the radio; typically only visible at kpc distances during flares \citep[e.g.,][]{Osten00}.
One abnormal radio-emitting binary observed in the GC M10 with a low measured mass function is possibly a RS CVn type of AB, or an unusual CV, or a face-on BH LMXB \citep{Shishkovsky18}. 

It is then intriguing to investigate if there exists a dependence, similar to that revealed by X-ray observations, of radio source populations on GC dynamical parameters. %Particularly, 
In this work, we %focus on
compare radio source counts of different GCs at the same luminosity cutoff, %constraining the significance of potential
searching for correlations between the number of radio sources vs. stellar encounter rate ($\Gamma$) and/or GC mass ($M$). The paper is organized as follows: in \S \ref{sec:observations_and_methods}, we describe the MAVERIC observations involved in this work, presenting the method we used to fit the data; in \S \ref{sec:results} we present results and discussions based on the results, and in \S \ref{sec:discussion_and_conclusion}, we draw conclusions.

% termed X-ray binaries (XRBs), where matter is accreted from an ordinary star to a compact object (neutron star or black hole), releasing gravitational energy, manifested as bright X-ray sources with typical X-ray luminosity of $10^{36-38}~\mathrm{erg~s^{-1}}$ \citep{Giacconi74, Clark75a, Canizares75}. Fainter sources with $L_X \lesssim 10^{35}~\mathrm{erg~s^{-1}}$ were also discovered, first with the {\it Einstein Observatory} \citep{Hertz83}, and later with {\it ROSAT}. It was not until the advent of {\it Chandra X-ray Observatory}, with high sensitivity and superior angular resolution, that the number of discovered GC X-ray sources quickly grew.

% Since the launch of early missions like {\it Uhuru} and {\it OSO-7}, it has been found that globular clusters (GCs) host an overabundance of X-ray sources. The number of X-ray sources per unit mass in GCs is orders of magnitude higher than that of the Galactic field \citep{Clark75b}. 

\section{Observations and Method}
% The sample of GCs observed by the MAVERIC survey are mostly massive ($\gtrsim 10^{5} M_\sun$) and nearby ($<9~\mathrm{kpc}$), to maximize sensitivities. There are $25$ clusters surveyed by {\it VLA} while $26$ by {\it ATCA},
\label{sec:observations_and_methods}
We use the {\it VLA} catalog of $5~\sigma$ radio point sources from the MAVERIC survey (Sh20). The catalog compiles radio point sources for 25 GCs, reporting source coordinates and radio flux densities in low and high frequency bands at $5~\mathrm{GHz}$ ($S_\mathrm{low}$) and $7.2~\mathrm{GHz}$ ($S_\mathrm{high}$)\footnote{Note that these frequencies are average central frequencies; the actual values may be slightly different between clusters.}, respectively. For each GC, we count the numbers of radio sources within the core radius ($r_c$) and the half-light radius ($r_h$). A radio source is selected if its $5~\mathrm{GHz}$ radio luminosity ($L_\mathrm{low}$) at the host cluster distance is higher than a limiting luminosity ($L_\mathrm{lim}$). We set $L_\mathrm{lim}=5.0\times 10^{27}~\mathrm{erg~s^{-1}}$ to include most relatively bright radio sources, %while ruling out 
excluding only the 
distant GCs M2, M3, and M54, as $L_\mathrm{lim}$ converts to a limiting flux ($S_\mathrm{lim}$) below the $<5\sigma$ detection limit of the catalog for those clusters. The source counts and relevant GC parameters are summarized in Table \ref{tab:radio_source_counts}. 

% \begin{table*}
%     \centering
%     \caption{Radio source counts in differnt GCs.}
%     \begin{tabular}{ccccccccccc}
%     \hline
%     \hline
%       GC Name & Distance & $r_c$ & $r_h$ & $S_\mathrm{lim}$ & $N_\mathrm{core}$ & $N_\mathrm{half}$ & $\mu_\mathrm{b, core}$ & $\mu_\mathrm{b, half}$ & $\Gamma$ & Mass \\
%         & ($\mathrm{kpc}$) & ($\arcmin$) & ($\arcmin$) & $\mathrm{\mu Jy}$ & & & & & & $\times 10^5~M_\sun$ \\
%     \hline
%     M4 & $1.8$ & $1.16$ & $4.33$ & $245.5$ & $0$ & $2$ & $0.1$ & $1.5$ & $27$ & $0.9$ \\
%     M5 & $7.7$ & $0.44$ & $1.77$ & $13.4$  & $0$ & $2$ & $0.4$ & $6.7$ & $164$ & $3.7$ \\
%     \end{tabular}
%     \label{tab:radio_source_counts}
% \end{table*}

\begin{deluxetable*}{lCCCCCCCCCCCC}[htb!]
%\tablenum{1}
\tablewidth{\textwidth}
\tablecaption{Radio source counts in different GCs\label{tab:radio_source_counts}}
\tablewidth{0pt}
\tablehead{
\colhead{GC Name} & \colhead{Distance} & \colhead{$r_c$} & \colhead{$r_h$} & \colhead{$S_\mathrm{lim}$} & \colhead{$N_\mathrm{core}$} & \colhead{$N_\mathrm{half}$} & \colhead{$N_\mathrm{m, core}$} & \colhead{$N_\mathrm{m, half}$} & \colhead{$\mu_{b, \mathrm{core}}$} & \colhead{$\mu_{b, \mathrm{half}}$} & \colhead{$\Gamma$} & \colhead{$M$} \\
\colhead{} & \colhead{(kpc)} & \colhead{(\arcmin)} & \colhead{(\arcmin)} & \colhead{($\mathrm{\mu Jy}$)} & \colhead{} & \colhead{} & \colhead{} & \colhead{} & \colhead{} & \colhead{} & \colhead{} & \colhead{($\times 10^5 M_\sun$)}
}
% \colnumbers
\startdata
M62$^\ast$       & 6.7 & 0.22 & 0.92 & 18.6  & 1 & 3  & 1 & 2 & 0.07 & 1.16 & 1670^{+710}_{-570} & 6.90 \pm 0.05 \\
NGC 6440$^\ast$  & 8.5 & 0.14 & 0.48 & 11.6  & 1 & 2  & 1 & 1 & 0.07 & 0.78 & 1400^{+630}_{-477} & 5.33 \pm 0.65 \\
M28$^\ast$       & 5.5 & 0.24 & 1.97 & 27.6  & 1 & 4  & 1 & 1 & 0.05 & 3.48 & 648^{+85}_{-91}    & 2.84 \pm 0.15 \\
M30              & 8.6 & 0.06 & 1.03 & 11.3  & 0 & 1  & 0 & 0 & 0.01 & 3.58 & 324^{+124}_{-81}   & 1.39 \pm 0.06 \\
M92              & 8.9 & 0.26 & 1.02 & 10.6  & 0 & 3  & 0 & 0 & 0.23 & 3.51 & 270^{+30}_{-29}    & 3.12 \pm 0.04 \\
M19              & 8.2 & 0.43 & 1.32 & 12.4  & 1 & 3  & 0 & 0 & 0.40 & 3.74 & 200^{+67}_{-39}    & 6.57 \pm 0.33 \\
M5               & 7.7 & 0.44 & 1.77 & 14.1  & 0 & 2  & 0 & 0 & 0.42 & 6.73 & 164^{+39}_{-30}    & 3.68 \pm 0.06 \\
M9               & 7.8 & 0.45 & 0.96 & 13.7  & 1 & 1  & 0 & 0 & 0.43 & 1.98 & 131^{+59}_{-42}    & 3.21 \pm 0.25 \\
M14              & 9.3 & 0.79 & 1.30 & 9.7   & 4 & 14 & 0 & 0 & 2.11 & 5.71 & 124^{+32}_{-30}    & 7.39 \pm 0.37 \\
NGC 6304         & 5.9 & 0.21 & 1.42 & 24.0  & 0 & 1  & 0 & 0 & 0.04 & 1.81 & 123^{+54}_{-22}    & 2.11 \pm 0.18 \\
NGC 6325         & 6.5 & 0.03 & 0.63 & 19.8  & 0 & 0  & 0 & 0 & 0.001& 0.54 & 118^{+45}_{-46}    & 0.76 \pm 0.13 \\
NGC 6544         & 3.0 & 0.05 & 1.21 & 92.9  & 0 & 0  & 0 & 0 & 0.00 & 0.33 & 111^{+68}_{-36}    & 1.15 \pm 0.11 \\
M22              & 3.1 & 1.33 & 3.36 & 86.9  & 0 & 1  & 0 & 0 & 0.40 & 2.54 & 78^{+31}_{-26}     & 4.09 \pm 0.04 \\
M13              & 7.6 & 0.62 & 1.69 & 14.5  & 1 & 5  & 1 & 1 & 0.83 & 6.13 & 69^{+18}_{-15}     & 4.69 \pm 0.20 \\
NGC 6760         & 7.4 & 0.34 & 1.27 & 15.3  & 0 & 1  & 0 & 0 & 0.25 & 3.46 & 57^{+27}_{-19}     & 2.55 \pm 0.30 \\
NGC 6539$^\ast$  & 7.8 & 0.38 & 1.70 & 13.7  & 2 & 5  & 1 & 1 & 0.04 & 6.20 & 42^{+29}_{-15}     & 2.61 \pm 0.31 \\
M10              & 4.4 & 0.77 & 1.95 & 43.1  & 0 & 3  & 0 & 0 & 0.25 & 1.59 & 31^{+5}_{-4}       & 1.89 \pm 0.04 \\
NGC 6712         & 8.0 & 0.76 & 1.33 & 13.1  & 2 & 2  & 1 & 1 & 1.24 & 3.80 & 31^{+5}_{-7}       & 1.19 \pm 0.08 \\
M4               & 1.8 & 1.16 & 4.33 & 258.0 & 0 & 2  & 0 & 0 & 0.11 & 1.49 & 27^{+12}_{-10}     & 0.93 \pm 0.02 \\
M12              & 5.2 & 0.79 & 1.77 & 30.9  & 0 & 1  & 0 & 0 & 0.38 & 1.93 & 13^{+5}_{-4}       & 0.87 \pm 0.03 \\
M107             & 6.1 & 0.56 & 1.73 & 22.5  & 0 & 1  & 0 & 0 & 0.28 & 2.69 & 6.8^{+2.3}_{-1.7}  & 0.81 \pm 0.05 \\
M55              & 5.7 & 1.80 & 2.83 & 25.7  & 2 & 5  & 0 & 0 & 2.91 & 7.19 & 3.2^{+1.4}_{-1.0}  & 1.88 \pm 0.07 \\
\enddata
\tablecomments{The distances are from \citet{Tremou18}; core radii ($r_c$), half-light radii ($r_h$) are from \cite{Harris96} (2010 edition); the $\Gamma$'s are from \cite{Bahramian13}; the GC masses ($M$) are from \cite{Baumgradt17} (2nd version)\footnote{\url{https://people.smp.uq.edu.au/HolgerBaumgardt/globular/}}. $N_\mathrm{core}$ and $N_\mathrm{half}$ are the observed numbers of radio sources within the core and the half-light region, with $L>5\times 10^{27}~\mathrm{erg~s^{-1}}$. $N_\mathrm{m, core}$ and $N_\mathrm{m, half}$ are numbers of confirmed members within the core radius and the half-light radius. GCs indicated with $^\ast$ have a significant radio source excess in the core over the expected background.}
\end{deluxetable*}

To fit the data, we follow the maximum likelihood method described in \citet[][V08 hereafter]{Verbunt08}. %For a complete view, 
We briefly outline the method here. 

The number of radio sources observed within the core or half-light radius of a GC follows a Poisson distribution: 
\begin{equation}
    P\lrb{N, \mu} = \frac{\mu^N}{N!}e^{-\mu},
    \label{eq:Poisson_distribution}
\end{equation}
where $\mu$ is the expected number of sources, and $N$ is the observed number of sources. This formula applies to both the number of actual cluster members ($N_c$) and the number of background sources ($N_b$), with $\mu_c$ members or $\mu_b$ background sources expected\footnote{Note that here $\mu_c$ and $\mu_b$ are shorthands for the expected numbers of members and background sources; in our application, they can represent the numbers in the core (indicated by a ``core" subscript) or within the half-light region (indicated by a ``half" subscript) as listed in Table \ref{tab:radio_source_counts}.}. $\mu_b$ is calculated 
for each cluster using
the normalized source counts from Sh20 (see table 4 of Sh20), given in  differential form $S_\nu^{2.5}dN/dS_\nu$, while applying the $S_\mathrm{lim}$ of each cluster as the lower bound of our integration. We assume that the expected number of cluster members, $\mu_c$, is %described
determined 
by  
$\Gamma$ and cluster mass ($M$). Given the low %source counts 
numbers of sources 
(Table \ref{tab:radio_source_counts}) at our designated luminosity cutoff ($L_\mathrm{lim}$), we follow a simplified linear model as in V08, viz.
\begin{equation}
    \mu_c = a\Gamma + b M,
    \label{eq:mu_c}
\end{equation}
where $a$ and $b$ are positive coefficients, while for convenience, we re-normalized $\Gamma$ and $M$, taken from Table \ref{tab:radio_source_counts}, to fractions of the $\Gamma$ of M62 and the $M$ of M14. \coryfr{The form of the model is based on the assumption that one expects more LMXBs and MSPs in GCs with higher $\Gamma$ and/or $M$. The positive correlation with $\Gamma$ has been tested by using census of X-ray sources from archival observations \citep[e.g.,][]{Pooley06} and radio timing surveys \citep{Hui10}.}

The likelihood function is then the multiplication of the joint probability, $P\lrb{N_c, \mu_c}P\lrb{N_b, \mu_b}$, over all GCs in our sample, i.e.,
\begin{equation}
    \mathcal{L} = \prod_i P\lrb{N_{c,i}, \mu_{c,i}}P\lrb{N_{b,i}, \mu_{b,i}},
\end{equation}
where $i$ indexes the GCs. The best-fitting model is given by a combination of $a$ and $b$ that maximizes $\mathcal{L}$.

\coryfr{We set up a grid of $a$ and $b$ values ranging from 0 to 4, with a spacing of $0.02$ in both $a$ and $b$, and for each pair of $a$ and $b$, we generate $1000$ random Poisson realizations.} Each realization draws a random integer as per a Poisson distribution given $\mu_c$, which is assigned to $N_c$; $N_b$ is then calculated by subtracting $N_c$ from the observed number of radio sources as listed in the 6th and the 7th column of Table \ref{tab:radio_source_counts}. \coryfr{To use the Poisson probabilities, we keep $N_c$ below the observed number of sources ($N_\mathrm{core}$ or $N_\mathrm{half}$), avoiding negative $N_b$ values; this is done by setting probability $P(N_b, \mu_b)=0$ when $N_b$ is negative, while keeping the total number of realizations ($1000$) for each pair of $a$ and $b$ unchanged. Zero probability leads to zero likelihood, so equivalently these realizations are excluded from the maximization of the combined likelihood ($\mathcal{L}$).}

\coryfr{We also perform a somewhat more constrained fit by keeping $N_{c}$ to be at least the number of confirmed members in the cores ($N_\mathrm{m, core}$, the 8th column in Table \ref{tab:radio_source_counts}) or half-light regions ($N_\mathrm{m, half}$, the 9th column of Table \ref{tab:radio_source_counts}). The members include confirmed core MSPs and LMXBs listed in Table \ref{tab:core_sources}, and PSR J1701$-$3006A in M62, which is outside  the core \citep{Lynch12}. Similarly, realizations with $N_{c}<N_\mathrm{m, core}$ or $N_\mathrm{m, half}$ are excluded from the fit by setting $P(N_c, \mu_c)=0$. For simplicity, this fit is referred to as the ``constrained fit"---to differentiate it from the ``unconstrained fit" where we only apply upper bounds on $N_c$ ($N_c$ less than or equal to the observed number of sources).}

\coryfr{To plot the result and compute confidence contours, we follow the definition of $Z$ in V08 that
\begin{equation}
    Z\equiv-2\lrbs{\log\lrb{\mathcal{L}} - \log\lrb{\mathcal{L}}_\mathrm{max}}.
    \label{eq:Z}
\end{equation}
Confidence contours are calculated assuming that $Z$ follows a $\chi^2$ distribution with one degree of freedom, so the best fit ($\mathcal{L}=\mathcal{L}_\mathrm{max}$) corresponds to $Z=0$, while the $68\%$, $90\%$, and $99\%$ confidence intervals are where $\Delta Z = 1.00$, $2.71$, and $6.63$, respectively.}
%We set $P\lrb{N_b, \mu_b} = 0$ if this happens. we noted that some GCs host zero or small numbers of sources in the core, which can result in negative $N_b$ values; we set the probability $P(N_b, \mu_b)$ to zero for these cases. 

% \begin{equation}
%     L = \prod_{i}P\lrb{N_{c,i}, \mu_{c,i}}P\lrb{N_{b,i}, \mu_{b,i}}.
%     \label{eq:likelihood}
% \end{equation}
% Given the low source counts.

\section{Results and Discussion}
\label{sec:results}
% \subsection{Results}
% We start by trying a simple model where $\mu_c$ is only dependent on one parameter ($\Gamma$ or $M$), i.e., $a=0$ or $b=0$. When $b$ is fixed to zero, the best fit gives $a=1.19$, with a $90\%$ confidence interval of $[0.16, 3.42]$ for sources in the core, while $a=0$ for source counts in the half-light region. The best-fitting $b=0.78$ when $a$ is fixed to zero, with a $90\%$ confidence interval of $[0.16, 2.21]$ for sources in the core; similarly, $b=0$ when the model fit to sources in the half-light region. Following V08, the confidence intervals can be calculated assuming that 
% \begin{equation}
%     Z\equiv-2\lrbs{\log\lrb{\mathcal{L}} - \log\lrb{\mathcal{L}}_\mathrm{max}}
%     \label{eq:Z}
% \end{equation}
% follows a $\chi^2$ distribution with one degree of freedom. The best fit then corresponds to $\mathcal{L}=\mathcal{L}_\mathrm{max}$, so $Z=0$, while the $68\%$, $90\%$, and $99\%$ confidence intervals are where $\Delta Z = 1.00$, $2.71$, and $6.63$ \citep{Avni76}, respectively. 

%In Figure \ref{fig:fit_a_only}, we plot $Z$ for either case, together with confidence intervals.

\begin{figure*}[htb!]
    \centering
    \includegraphics[width=\textwidth]{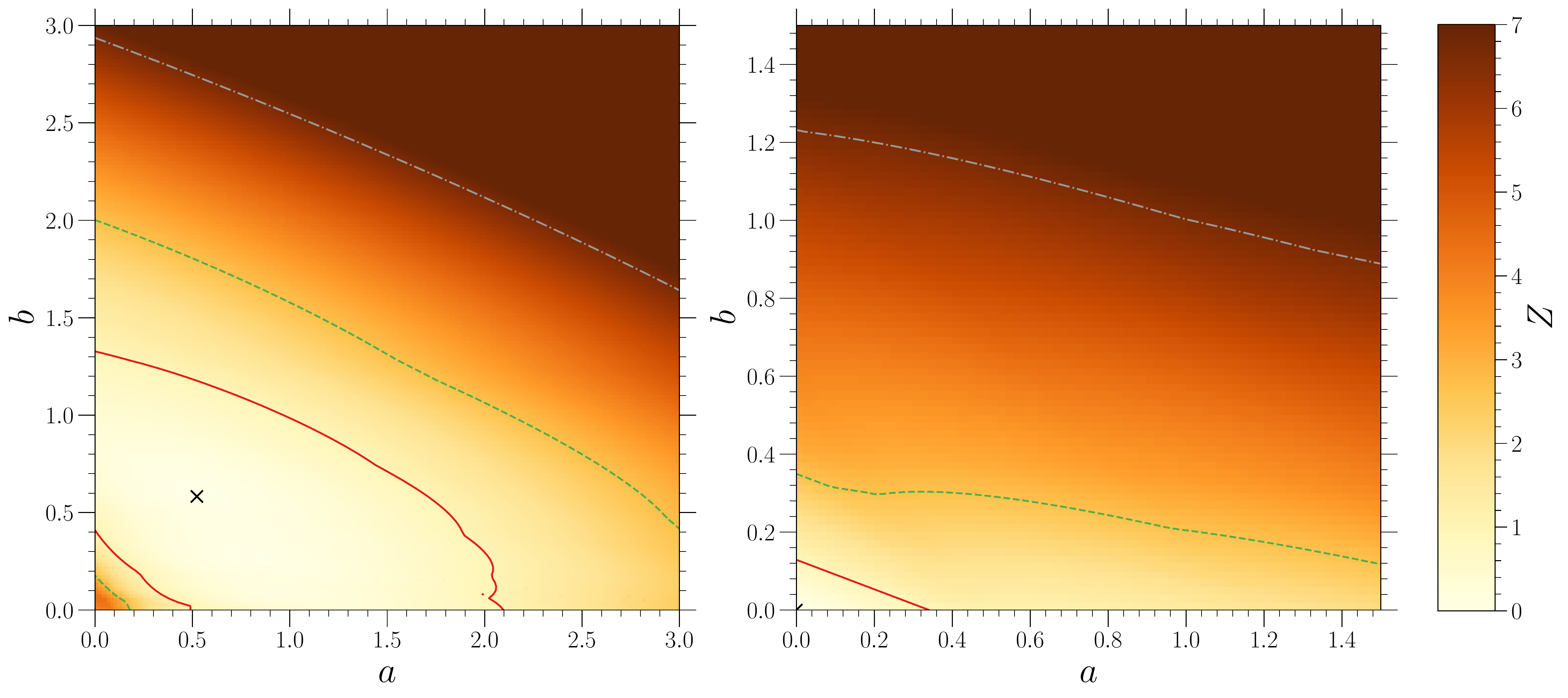}
    \caption{$68\%$ (solid red), $90\%$ (dashed green), and $99\%$ (dashed-dotted grey) contours in the $ab$ plane for fits to radio source counts in cluster cores (left) and half-light regions (right). The best-fitting values of $a$ and $b$ are indicated with a black cross in each panel. The colorbar presents intensity of $Z$. For source counts in the core, $a \neq 0$ or $b \neq 0$ at the $90\%$ level, while $a=b=0$ %at $90\%$ confidence 
    is consistent with the data for sources within the  half-light regions. %i.e., no clear dependence of source counts on $\Gamma$ and/or $M$.
    }
    \label{fig:contour_plots}
\end{figure*}

\begin{figure*}
    \centering
    \includegraphics[width=\textwidth]{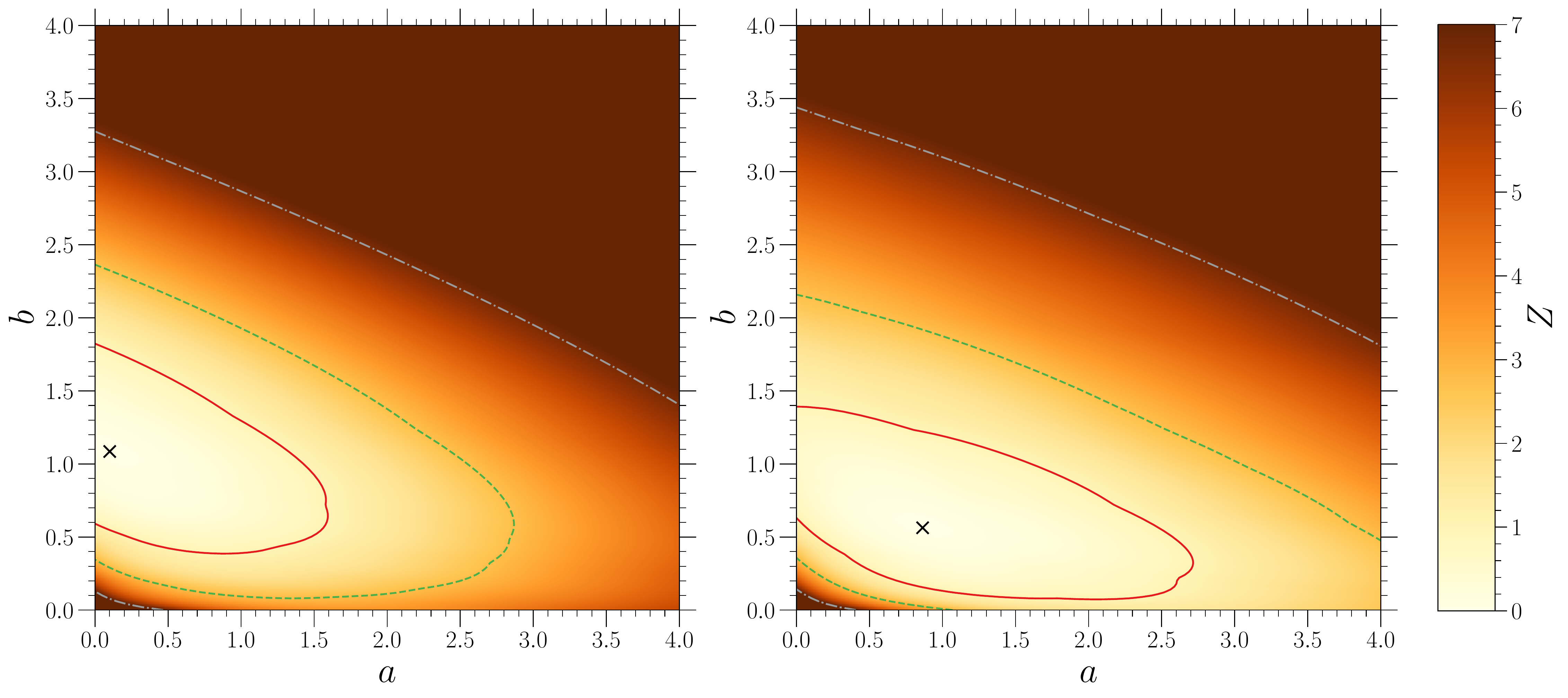}
    \caption{The marker and the contours are the same as those of Figure \ref{fig:contour_plots}, but this fit required  $N_c$ to be equal to or greater than the number of confirmed members in the core (left) or within the half-light region (right).}
    \label{fig:fit_with_lower_limit}
\end{figure*}

\coryfr{The resulting distributions of $Z$ for unconstrained and constrained fits are presented in Figure \ref{fig:contour_plots} and \ref{fig:fit_with_lower_limit}, overplotted with confidence contours. The unconstrained fit to source counts in the core gives the best-fitting $a=0.52$ and $b=0.58$, excluding non-correlation ($a=b=0$) at the $90\%$ confidence level; whereas the best fit to source counts within the half-light region suggests no correlation: $a=b=0$. }

\coryfr{The constrained fit to source counts in the core consistently favors a correlation (best fit: $a=0.10$ and $b=1.09$), %but 
now 
at the $99\%$ confidence level. Furthermore, the entire $a$-axis lies outside the 90\% confidence contour,
%suggesting
indicating 
a dependence of source counts on GC mass ($b\neq 0$) at $90\%$ confidence. We note that the constrained fit to source counts within the half-light region (best fit: $a=0.86$, and $b=0.56$) also excludes non-correlation at $99\%$ confidence, in contrast to the unconstrained fit; while it rules out the $\Gamma$-only dependence ($b=0$) at a lower ($68\%$) confidence level.}

\coryfr{For both the constrained and unconstrained fits, we generate $1000$ random data sets based on the best-fitting $a$ and $b$ and calculate the corresponding combined likelihood $\mathcal{L}$ for each data set. We note that the maximum likelihoods corresponding to the best fits are greater than those for all random data sets; the best fits are therefore appropriate for modelling the radio source counts.}

% We start by investigating fits where $a$ and $b$ to vary and find the best-fitting $a=0.52$, and $b=0.58$ for source counts in the core. The best fit results in $a=b=0$ for sources in the half-light region. In Figure \ref{fig:contour_plots}, we show the parameter planes of $a$ and $b$, and the confidence contours. 

% The resulting $Z$ values are plotted in Figure \ref{fig:fit_with_lower_limit}. \coryfr{For core source counts,} the best-fitting $a=0.10$ and $b=1.09$; the origin ($a=b=0$) lies outside the $99\%$ confidence contour, while the entire $a$-axis lies outside the 90\% confidence contour. \coryfr{We note that the constrained fits give non-zero best-fitting parameters: $a=0.86$, and $b=0.56$, for source counts in half-light regions, in contrast to the unconstrained fit (Figure \ref{fig:contour_plots}). The following discussions will focus on the constrained fits.}

% \begin{figure*}[htb!]
%     \centering
%     \includegraphics[width=\textwidth]{fit_single_parameter.pdf}
%     \caption{$Z$ (defined in eq.\ref{eq:Z}) plotted for the case where $\mu_c$ is assumed to only depend on $\Gamma$ (left panels), and where $\mu_c$ is assumed to only depend on $M$ (right panels). Top panels plot $Z$ for fits to source counts in the core, while the bottom panels plot $Z$ for fits to source counts in the half-light radius. %In all panels, 
%     The best fitting $a$ or $b$ is indicated with a vertical gray line, while the $68\%$ and $90\%$ confidence intervals are marked by dashed red lines and dash-dotted green lines, respectively. Note that, since the best $a$ or $b$ is zero in the half-light radius, only upper bounds of the confidence intervals are plotted for the lower panels.}
%     \label{fig:fit_a_only}
% \end{figure*}

%Thus, this fit indicates that the number of radio sources does not depend exclusively  on $\Gamma$, at 90\% confidence, and can be explained with a dependence on $M$ alone or both $M$ and $\Gamma$.
% \subsection{Discussion}
% Our results suggest that the radio source counts in the core do depend on $\Gamma$ and/or $M$ at $90\%$ confidence level, \coryfr{or at $99\%$ level when confident members are considered. The latter is more constrained and indicates that radio source counts in the core does not depend exclusively on $\Gamma$ at $90\%$ confidence (Figure \ref{fig:fit_with_lower_limit}).} 

\begin{figure*}[!htb]
    \centering
    \includegraphics[width=\textwidth]{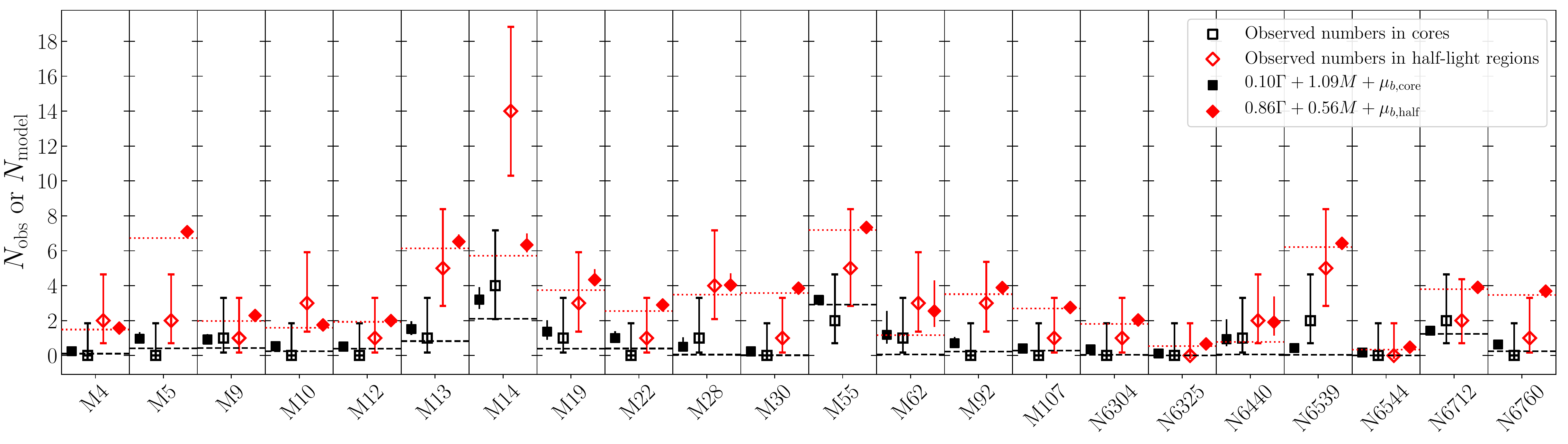}
    \caption{\coryfr{Observed core source counts (black empty squares) and observed half-light region source counts (red empty diamonds) vs. predicted source counts in the core (black filled squares) and in the half-light region (red filled diamonds) based on the constrained fits plotted for all 22 GCs. The error bars on the observed numbers indicate $84\%$ confidence upper and lower limits \citep{Gehrels86}; for GCs with zero counts, we only plot the upper bar. The uncertainties in predicted numbers are propagated from uncertainties in $\Gamma$ and $M$, and the $68\%$ confidence ranges of $a$ and $b$. The black dashed and red dotted lines in each panel marks the number of expected background sources in the core ($\mu_{b,\mathrm{core}}$) and in the half-light region ($\mu_{b,\mathrm{half}}$), respectively.}}
    \label{fig:compare}
\end{figure*}

Looking at Table \ref{tab:radio_source_counts}, we note that a few GCs (indicated with an $\ast$) show significant core source excesses over the predicted numbers of background sources (6th column vs. 10th column). Statistically, the probability of observing $1$ while expecting $0.04-0.07$ background sources (Table \ref{tab:radio_source_counts}) is around $4\%-7\%$ (eq. \ref{eq:Poisson_distribution}), so the excess is not very likely given by background sources. \cory{In Figure \ref{fig:compare}, we plot the model predicted counts using the best-fitting parameters vs. the observed counts in the core for each GC. We also overplot error bars that indicate uncertainties propagated from the fitting parameters ($a$ and/or $b$) and from the corresponding GC parameters ($\Gamma$ and/or $M$). %GCs indicated with $\ast$ in Table \ref{tab:radio_source_counts} possess significantly more observed counts than those predicted by best-fitting models. Such discrepancies, however, would be mitigated if uncertainties in $a$ and $b$ are included.
}

Source counts within the half-light region have no clear dependence on either $\Gamma$ or $M$ (with the best-fitting $a$ and $b=0$), when $N_c$ is only limited by $N_\mathrm{half}$. \coryfr{However, when confirmed members are specified, the result excludes non-dependence ($a=b=0$) at $99\%$ confidence. Of the $6$ GCs that have confirmed members, three (M62, NGC 6440, and M28) are large in both $\Gamma$ and $M$ (Table \ref{tab:radio_source_counts}); specifying confirmed members for these GCs favors a positive correlation between $\mu_c$ and $\Gamma$ and/or $M$, although the observed source counts ($N_\mathrm{half}$) are generally consistent with the predicted numbers of background sources ($\mu_{b, \mathrm{half}}$), within the $90\%$ confidence level derived for Poisson statistics \citep[Gehrels upper and lower limits; see][]{Gehrels86}.}

% One might argue the observed source counts ($N_\mathrm{half}$) are generally consistent with the predicted numbers of background sources ($\mu_{b, \mathrm{half}}$), within the $90\%$ confidence level derived for Poisson statistics \citep{Gehrels86}; this, however, does not contradict the fit result because specifying confirmed members is equivalent to endowing GCs with source excess over the background estimate; moreover, the $6$ GCs that have confirmed members are those large in $\Gamma$ and/or $M$ (Table \ref{tab:radio_source_counts})

\coryfr{One exception is M14---the observed half-light region source count ($14$) is consistent with the background estimate ($5.71$) at a higher, $98\%$, confidence level, considering the Gehrels upper limit. We note that M14 has the lowest $S_\mathrm{lim}$ of $\approx 9.2~\mathrm{\mu Jy}$, and for all GCs but M14, $S_\mathrm{lim}$ is above the very first flux bin of Sh20's radio source counts (between $7.90~\mathrm{\mu Jy}$ and $10.61~\mathrm{\mu Jy}$; see table 4 of Sh20). For M14, part of this bin goes below the $5~\sigma$ flux limit of the observation ($\approx 9~\mathrm{\mu Jy}$). Including this flux bin in the integration will incorporate faint sources, between $9$ and $10.61~\mathrm{\mu Jy}$, so gives a somewhat higher background estimate ($7.82$) for M14. However, the first flux bin also misses faint sources between $7.9~\mathrm{\mu Jy}$ and $9~\mathrm{\mu Jy}$. It is hence clear that the expected number of background sources ($\mu_{b,\mathrm{half}}$) might have been underestimated for M14 even when the first flux bin is considered. We run our constrained fits without M14 and note that the best fit to core radio source counts consistently excludes $a=b=0$ at $99\%$ confidence, with the maximum likelihood at $a=0.08$ and $b=1.03$; and the best fit to the half-light region counts also excludes $a=b=0$ at $99\%$ confidence, giving $a=0.93$ and $b=0.46$.}

% ($N_\mathrm{half}=14$ while $\mu_{b, \mathrm{half}} = 5.71$, consistent at $95\%$ confidence level)
% If we include the first flux bin in the integration for M14, the first bin is included in the integration. considering its lower bound ($7.78~\mathrm{\mu Jy}$) is below the $5~\sigma$ limit of most GCs in the list.%and it is consistent with 

The $22$ GCs listed in Table \ref{tab:radio_source_counts} host a total of 16 sources in their cores. We further investigate all these sources by matching their radio positions to those measured by timing observations (Table \ref{tab:core_sources}, and references therein). We find $4$ known MSPs, three of which are in M28, NGC 6440, and NGC 6539, GCs that have core source excesses over the background (Table \ref{tab:radio_source_counts}), and one in M13, which is marginally above $L_\mathrm{lim}$ (Table \ref{tab:core_sources}). %From multiple references, we also found 
Two more sources are known cluster members; a BH candidate in M62 \citep{Chomiuk13} and  %an ultra-compact X-ray binary (UCXB) 
a NS LMXB 
in NGC 6712 \citep{Swank76}. 

These sources are all 
%more
confirmed cluster members, with $4$ hosted by GCs that have a core radio source excess, namely M28, M62, NGC 6440, and NGC 6539.  Three of these are GCs with high mass and $\Gamma$ (Table \ref{tab:radio_source_counts}), in agreement with the positive correlation shown by the fit. 
%Statistically, the probability of observing $1$ while expecting $0.04-0.07$ background sources (Table \ref{tab:radio_source_counts}) is around $4\%-7\%$ (eq. \ref{eq:Poisson_distribution}), so the excess is not very likely given by background sources.

%One likely outlier to the correlation is 
NGC 6539 
%, which also 
possesses a clear core source excess ($2$ observed vs. $0.04$ expected), while having a low $\Gamma$ of $42$ and an intermediate $M$ ($=2.6\times 10^5~M_\sun$ vs. median mass of $2.8\times 10^5~M_\sun$). There are two core sources, VLA11 and VLA12. VLA11 is positionally consistent with the known MSP, PSR B1745$-$20 \citep{DAmico93}, while VLA12 has an inverted-to-steep index  ($\alpha=-0.4\pm0.7$, Table. \ref{tab:core_sources}), which can overlap with multiple scenarios. The probability of  finding a background source in the core is $\approx 8\%$, not very likely but not ruled out.
%, so the core source excess is significant. 
%The positional match of VLA11 to PSR B1745$-$20, however, is not as satisfactory as other matches---the radio timing position measured by \citet{DAmico93} lies $\approx 0.7\arcsec$ southeast to the {\it VLA} position, which is $\approx 14$ times the size of the $1~\sigma$ {\it VLA} positional uncertainty ($\approx 0.05\arcsec$; Sh20). The deviation cannot be fully accounted for by proper motion \citep{Gaia18}, despite that the discovery is over $20$ years apart from the {\it VLA} observation ($\Delta \mathrm{Ra}$ and $\Delta \mathrm{Dec}$ $\approx -0.1\arcsec$). Nevertheless, the source excess would still be significant even VLA11 was a background source, as VLA12 is very unlikely a background source. 
Considering the low $\Gamma$, it is possible that VLA12 is a primordial cluster member. For example, short ($\sim$ hours) flares observed in some short-period RS CVn ABs can be detected 
in the radio at kpc distances \citep[e.g.,][]{Osten00}.

MSPs in cluster cores are expected to scale directly with $\Gamma$, as they are produced by LMXBs that scale with stellar encounter rate \citep[e.g.][]{Heinke03,Pooley06,Hui10,Bahramian13}. The presence of four confirmed MSPs among the cluster radio sources indicates that at least some cluster radio sources must be dynamically formed. However, our analysis including known cluster members requires a contribution from cluster mass, not just stellar encounter rate.
 %This is expected, since known cluster radi of which are thought to be dynamically formed, and timates of stellar encounter rate \citep[e.g.][]{Bahramian13}. 
 %The positive correlation between the number of radio sources in the core ($\mu_c$) and $\Gamma$ suggests that some core radio sources may have a dynamical origin.
It is therefore of great interest that the relative numbers of radio-emitting BH LMXBs are not predicted to scale directly with relative stellar encounter rates $\Gamma$ (which are calculated for visible stars, not the BH subsystems). 
%On the other hand, it is not clear how the relative numbers of BH LMXBs will vary among globular clusters, since this 
The numbers of these BH binaries 
are instead predicted to depend in complicated ways on the total number of BHs in each cluster, and on the interactions of these BH populations with other cluster stars \citep{Weatherford18,ArcaSedda18}. 
Thus, our work gives tentative support to the idea that a portion of the cluster radio sources are BH LMXBs, distributed in a complicated way among clusters, although we cannot rule out a contribution by other kinds of sources.

%the compact objects of which were brought to the core by mass segregation before they encounter with stars to form close binaries. 

\begin{figure}
    \centering
    \includegraphics[width=\columnwidth]{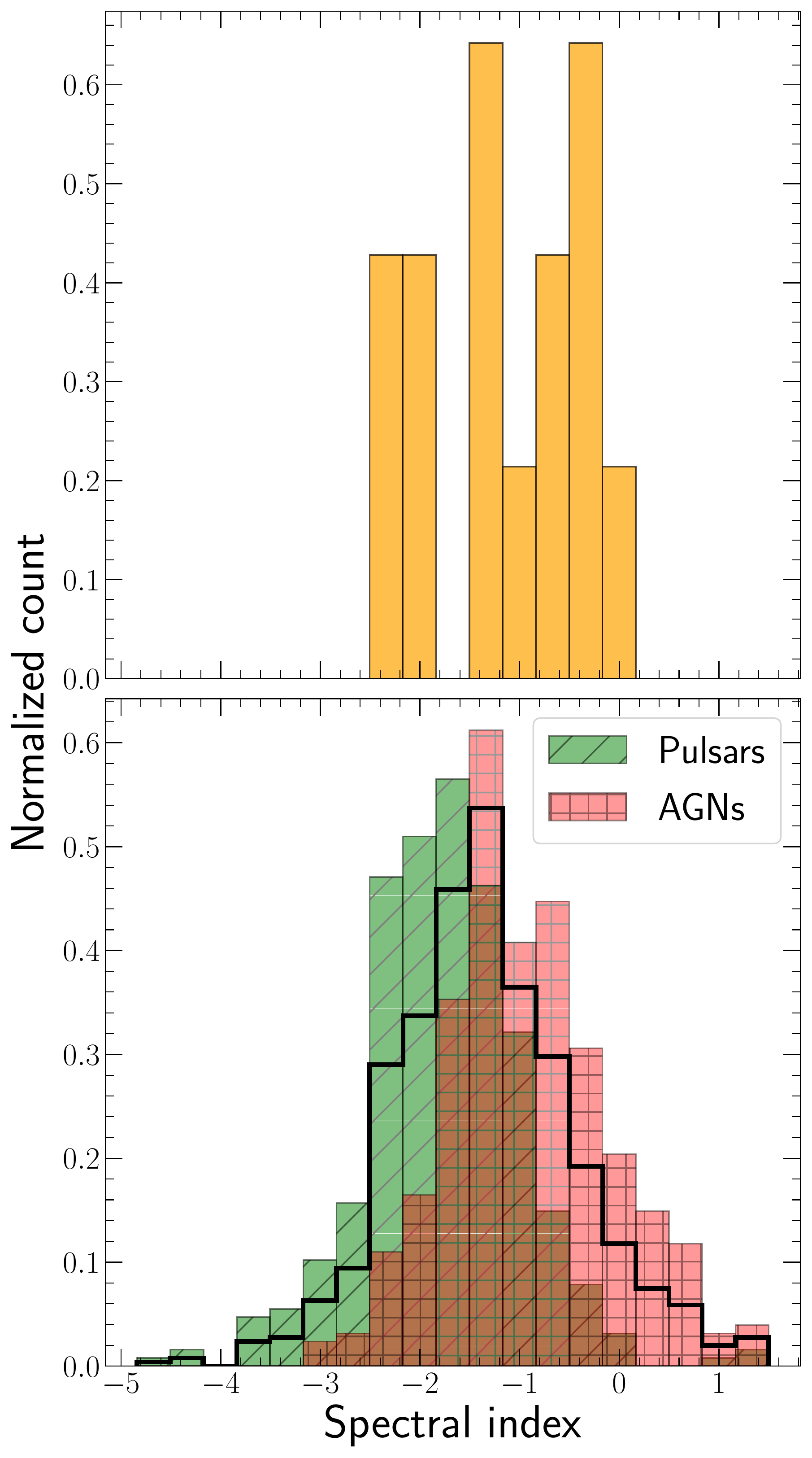}
    \caption{\coryfr{{\it Top}: A histogram showing distribution of spectral index ($\alpha$) for core radio sources with constrained $\alpha$ values as listed in Table \ref{tab:core_sources} (orange). {\it Bottom}: Histograms of $\alpha$ for pulsars from the ATNF catalog (hatched green), AGNs (hatched red), pulsar and AGN combined (solid black contour). The counts in each bin is normalized by dividing the total counts of each sample and the bin width. The core sources divide into steep-index and flatter-index groups; the flatter group is more numerous than expected from a pulsar-only distribution. Despite that the AGNs can partly account for the flatter group, there is an overabundance of core sources around $-0.5\lesssim \alpha \lesssim -0.2$.}}
    \label{fig:SI_comparison}
\end{figure}

\cory{%The source excesses can be contributed by both LMXBs (typically BH systems) and MSPs. 
Another interesting test involves comparing the radio spectral indices of our sample with the known distribution of spectral indices of MSPs. 
In Figure \ref{fig:SI_comparison}, we present normalized histograms of spectral indices ($\alpha$) for pulsars from the ATNF catalog \citep[][and references therein]{Manchester05}; sources from the MAVERIC catalog that are outside of the half-light radii and thus are mostly background active galactic nuclei (AGNs); and for radio sources in Table \ref{tab:core_sources} (excluding sources with unconstrained $\alpha$ values). \coryfr{We first compare equal-size samples of MSPs ($P\leq 20~\mathrm{ms}$) and normal pulsars from the ATNF catalog, and note that there is no significant difference between the distribution of their spectral indices (p-value of $0.25$ for a two-sample Anderson-Darling test). We can therefore} randomly draw pulsars from the whole ATNF catalog to form a sample the same size as that of the AGNs. The core sources seem to follow a bimodal distribution, containing a group of steep ($-2.5 \lesssim \alpha \lesssim-1.2$) sources, and a group of relatively flatter ($-0.8\lesssim\alpha\lesssim0.2$) sources. The latter significantly deviates from the observed pulsar distribution, and exceeds the AGN distribution around $-0.5\lesssim \alpha \lesssim -0.2$, which could be partially contributed by LMXBs. We performed a 2-sample Anderson-Darling test \citep{Scholz87} comparing the core source distribution with that of the ATNF pulsars and that of AGNs. A test comparing the core sources with ATNF pulsars rejects the null hypothesis that the core sources and  ATNF pulsars are drawn from the same population at a level more significant than $0.1\%$. A test comparing core sources with the AGN sample rejects the null hypothesis at $3\%$ significance. Finally, a test comparing the core sources with the AGN and pulsar samples combined rejects the null hypothesis at a level more significant than $0.1\%$. In fact, the flatter group contains $6$ sources, of which one is a NS LMXB, one is the known BH candidate in M62, and one is the known MSP in NGC 6539; the other 3 sources could be either background AGNs or cluster members.} 

% The lack of a clear detection of any correlation between the numbers of sources within the half-light radius and cluster properties could be the result of the dominance of background sources between the core and half-light radii. 

\coryfr{Our analyses have shown that additional membership information results in tighter constraints on the parameters for both source counts in the core and those in the half-light region.} In this regard, future careful multiwavelength follow-up will therefore be necessary to separate background sources from cluster members, allowing investigation of radial trends in the radio source properties. \cory{For example, proper motion analysis of potential optical counterparts, or with radio VLBI \citep{Tetarenko16}, can be effective in excluding background sources. Detailed investigation at other wavelengths is beyond the scope of this paper, but will be addressed %in a more systematic manner in future work of compilations of BH LMXB and MSP candidates.
in future work.} 

%Furthermore, given the heavy masses of binaries containing BHs, we expect them to mass segregate relatively quickly and be significantly centrally concentrated (e.g., as shown for NS LMXBs by \citealt{Grindlay84} and for MSPs by \citealt{Grindlay02}).

% \subsection{Caveats}
\cory{There are several %factors that could contribute uncertainties to the final results. 
potential factors that could affect our final results; uncertainties in $\Gamma$ or mass of clusters, differences in sensitivity to radio sources across the cluster cores, and unusual cluster histories.  
%In particular, 
Uncertainties in $\Gamma$ can be up to $70\%$ of the central values \citep{Bahramian13}, but are not included in our aforementioned fits. To address this, we fit $a$ and $b$ simultaneously to the same observed counts but with $\Gamma$ and $M$ randomly drawn from the error ranges for each cluster, and observe the distribution of the best-fitting values and the corresponding confidence contours. We ran a total of 50 random constrained fits and find the point $a=b=0$ is outside of the $99\%$ confidence contour in all of the fits; running $50$ random constrained fits for source counts in the half-light regions also excludes $a=b=0$ at $99\%$ confidence, so the correlation holds for both core and half-light region source counts, regardless of uncertainties in $\Gamma$ and $M$.}

% an average of best-fitting $a$ and $b$ of $0.65$ and $0.42$, with standard deviations of $0.20$ and $0.19$, respectively. The origin, $a=b=0$, is outside the $90\%$ confidence contour in all realizations, so the positive correlation with $a$ and/or $b$ is still at $90\%$ confidence even when including uncertainties in $\Gamma$ and $M$.

\cory{Another factor to consider is the increase in noise in the radio images with increasing off-axis distance. To first order, %one can assume that 
the RMS noise scales as the reciprocal of the primary beam sensitivity. As a result, sources at the edges of the beam might have a flux above the average $5~\sigma$ level, but might have been missed because they are below the local $5~\sigma$ level. This effect is expected to be minor on our analyses of the core sources, as most of the clusters have core sizes much smaller than the half-power primary beam size ($\approx 8.2\arcmin$ at $5.5~\mathrm{GHz}$). The largest core, in the cluster M55, is $1.8\arcmin$ in radius.
If we made the assumption that all of our core sources (Table \ref{tab:core_sources}) are located $1.8\arcmin$ from the beam center, %(which is the core size of M55, the largest among all GCs), 
we would expect the sensitivity to drop to $\approx 88\%$ of the center value (assuming the primary beam follows a Gaussian sensitivity curve); this corresponds to an increase in noise by a factor of $\approx 1.1$, so the detection procedure might %have a chance to 
miss sources between $5$ and $5.5~\mathrm{sigma}$. %There is 
Only one (of 16) of our core sources, M19-VLA34, falls in this flux range; we thus estimate that $<6\%$ of the core sources above our stated flux limit might have been missed by our observations. Each GC has 1--3 core sources, corresponding to 0.1--0.2 missing sources. Therefore, the effect of noise variation is minor on our analyses.

A final concern is whether unusual histories of these GCs could mean that the current mass and stellar interaction rate do not represent their values during the period in which X-ray binaries or MSPs were formed. The prototype for this argument is the GC NGC 6712, which was shown to have lost $>99\%$ of its mass, leaving preferentially high-mass stars and binaries in the core  \citep{DeMarchi99,Andreuzzi01}. This cluster was probably much more massive and dense when the two detected radio sources (one a known NS LMXB) were formed, explaining their presence in an apparently low-mass, low-density cluster \citep[see][]{Ferraro00}. However, running our fits excluding NGC 6712 gives best-fitting $a=0.32$ and $b=0.78$ \coryfr{for the core source counts, and $a=0.97$, $b=0.42$ for the half-light region counts}, while both fits consistently exclude $a=b=0$ at $99\%$ confidence (when lower limits on $N_c$ are included).
Although similar cluster mass loss is likely to have affected other clusters \citep[see e.g.][]{Moreno14}, this effect has not removed the detectable correlation of stellar interactions on X-ray sources \citep{Pooley06,Bahramian13}, and therefore we suspect its effects on radio sources are likely also modest. 
}

% and (2) increase in noise at larger radii. While the latter is not crucial for our analysis in the core, it could contribute to our underestimating sources in the half-light regions.

%Considering that core radio sources are mostly associated with relatively massive binaries like BH or NS LMXBs, or MSPs, the lack of excess in the half-light region could be the combined effect of relatively low rate of dynamical encounter and mass segregation. The former results in low formation rate of BH or NS LMXBs, while the later shift massive binaries or compact objects to the core so deplete massive binaries in the half-light region.
%The other sources with steep radio spectra could be potential MSPs. 

\begin{deluxetable*}{lCCCll}[htb!]
%\tablenum{1}
\tablecaption{Radio sources in the core with $L_\mathrm{low} > 5\times 10^{27}~\mathrm{erg~s^{-1}}$\label{tab:core_sources}}
\tablewidth{0pt}
\tablehead{
\colhead{Source ID} & \colhead{$L_\mathrm{low}$} & $S_\mathrm{low}/\mathrm{RMS}$ & $\alpha$ & Notes & References \\
\colhead{} & \colhead{($\times 10^{27}~\mathrm{erg~s^{-1}}$)} & \colhead{($\sigma$)} & \colhead{($S_\nu \propto \nu^\alpha$)} & \colhead{} & \colhead{}
}
% \decimalcolnumbers
\startdata
M9-VLA24       & 5.4  \pm 0.6 & 8.6  & -0.9^{+0.6}_{-0.6} & - & - \\
M13-VLA21      & 5.0  \pm 0.7 & 7.3  & -1.3^{+0.7}_{-0.8} & PSR B1639$+$36A & K91, W20 \\
M14-VLA8       & 24.2 \pm 1.0 & 26.1 & -1.9^{+0.3}_{-0.3} & - & - \\
M14-VLA11      & 20.0 \pm 1.0 & 21.4 & -1.2^{+0.3}_{-0.3} & - & - \\
M14-VLA15      & 18.6 \pm 1.0 & 20.0 & -0.2^{+0.2}_{-0.2} & - & - \\
M14-VLA45      & 6.3  \pm 1.0 & 6.8  & <0.0 & - & - \\
M19-VLA34      & 5.1  \pm 0.9 & 5.5  & <0.6 & - & - \\
M28-VLA3       & 24.3 \pm 0.5 & 53.7 & -2.2^{+0.1}_{-0.1} & PSR B1821$-$24A & L87, F88, C04, R04, B11, J13\\
M55-VLA6       & 8.7  \pm 0.4 & 19.6 & -1.2^{+0.3}_{-0.3} & - & - \\
M55-VLA15      & 5.1  \pm 0.5 & 11.3 & -2.3^{+0.7}_{-0.7} & - & - \\
M62-VLA1       & 6.0  \pm 1.0 & 7.0  & -0.4^{+0.6}_{-0.5} & A BH candidate  & C13 \\
NGC 6440-VLA6  & 33.2 \pm 1.3 & 28.4 & -2.0^{+0.2}_{-0.2} & PSR B1745$-$20  & L96, F08\\
NGC 6539-VLA11 & 9.4  \pm 0.7 & 9.4  & -0.7^{+0.7}_{-0.8} & PSR B1802$-$07  & D93, T93, T99 \\
NGC 6539-VLA12 & 5.2  \pm 0.6 & 8.5  & -0.4^{+0.7}_{-0.7} & - & - \\
NGC 6712-VLA7  & 33.8 \pm 0.9 & 38.4 &  0.1^{+0.1}_{-0.1} & 4U 1850$-$087 & S76, A93, H96, S06 \\
NGC 6712-VLA9  & 25.0 \pm 0.9 & 28.3 & -0.6^{+0.1}_{-0.1} & - & - \\
\enddata
\tablecomments{RMS noises are adapted from Sh20. $\alpha$ is the spectral index from Sh20. References: A93: \citet{Anderson93}, B11: \citet{Bogdanov11}, C04: \citet{Cognard04}, C13: \citet{Chomiuk13}, D93: \citet{DAmico93}, F88: \citet{Foster88}, F08: \citet{Freire08}, H96: \citet{Homer96}, J13: \citet{Johnson13}, K91: \citet{Kulkarni91}, L87: \citet{Lyne87}, L96: \citet{Lyne96}, R04: \citet{Rutledge04}, S76: \citet{Swank76}, S06: \citet{Sidoli06}, T93: \citet{Thorsett93}, T99: \citet{Thorsett99}, W20: \citet{Wang20}.}
\end{deluxetable*}

\section{Conclusion}
\label{sec:discussion_and_conclusion}
\coryfr{We investigate linear correlations of radio source counts with the encounter rate ($\Gamma$) and mass ($M$) for a total of 22 GCs from the MAVERIC survey. Including  information about confirmed cluster members in our analysis, the fit to source counts in the core rules out no correlation with $M$ or $\Gamma$ at 99\% confidence, and excludes a dependence on $\Gamma$ alone at 90\% confidence.  The fit to to source counts in the half-light region also rules out no correlation at $99\%$ confidence, but just excludes a  $\Gamma$-only dependence at 68\% confidence.}

The histogram of spectral indices of our radio sources are inconsistent with that of pulsars and/or AGN alone, indicating another component with flatter spectra is present.
Our findings are intriguingly consistent with the expectations that of detectable radio sources, some will be MSPs (produced dynamically, and thus correlated with $\Gamma$), and some may be BH LMXBs, the numbers of which are unlikely to be directly correlated with $\Gamma$ \citep{Kremer18,ArcaSedda18}.

\coryfr{Our analyses also suggest that membership information can lead to significantly tighter constraints on the parameters. Future secure determinations of the nature of individual systems, obtained with careful follow-up, will therefore allow us to test predictions of the distribution of BH binaries in globular clusters in more detail.} More detailed analyses will also use the full luminosity range of detected radio sources in each cluster, and (with sufficient follow-up efforts) will study their radial distributions. 

%The numbers of radio sources in the core and half-light region follow a positive correlation, at $99\%$ confidence level, with stellar encounter rate ($\Gamma$) and/or GC mass ($M$), 

%while no clear correlation was found for source counts in the half-light region (which we attribute to the preponderance of background AGN over cluster members in this region).
% Among the core radio sources, 5 are known to be MSPs or NS LMXBs (believed to correlate with stellar encounter rate), and one is a candidate BH binary. 

%We suggest that core radio sources are more likely LMXBs with NS or BH as the compact object, or MSPs, which are relatively more massive than general cluster population. %the lack of source excess in the half-light regions indicates the lack of these massive binaries, which could be the result of relatively low formation rate (as a result of low encounter rates), and/or mass segregation that have shifted the compact object and/or ever formed massive binaries to the core.

\section*{Acknowledgement}
COH \& GRS are supported by NSERC Discovery Grants RGPIN-2016-04602 and RGPIN-2016-06569, respectively. JS is supported by NSF grants AST-1308124 and AST-1514763, and a Packard Fellowship. JCAM-J is the recipient of an Australian Research Council Future Fellowship (FT140101082), funded by the Australian government. The MAVERIC source catalog used in this work is derived from observations made by the Karl G. Jansky Very Large Array under the National Radio Astronomy Observatory. The National Radio Astronomy Observatory is a facility of the National Science Foundation (NSF) operated under cooperative agreement by Associated Universities, Inc. (AUI).

\newpage

\bibliography{ref}{}
\bibliographystyle{aasjournal}
\end{document}